\documentclass[
 reprint,
superscriptaddress,
 amsmath,amssymb,
 aps,
]{revtex4-1}

\usepackage[utf8]{inputenc}
\usepackage{xcolor}
\usepackage{graphicx}

\begin{document}

\title{Hexagonal matching codes with 2-body measurements}
\author{James R. Wootton}
\affiliation{IBM Quantum -- IBM Research Zurich}
\date{\today}

\begin{abstract}
Matching codes are stabilizer codes based on Kitaev's honeycomb lattice model. The hexagonal form of these codes are particularly well-suited to the heavy-hexagon device layouts currently pursued in the hardware of IBM Quantum. Here we show how the stabilizers of the code can be measured solely through the 2-body measurements that are native to the architecture. The process is then run on 27 and 65 qubit devices, to compare results with simulations for a standard error model. It is found that the results correspond well to simulations where the noise strength is similar to that found in the benchmarking of the devices. The best devices show results consistent with a noise model with an error probability of around $1.5\%-2\%$.
\end{abstract}

\maketitle

\section{Introduction}

The future of quantum computation is built on quantum error correction~\cite{lidar:13}. It is therefore vital to design and test quantum hardware in a way that ensures compatibility with this process. It is for this reason that recent years have seen many examples of experiments testing the fundamental components of quantum error correction (see \cite{wootton:20,andersen:20,erhard:21,gong:21,chen:21,hilder:21,ryananderson:21} from the past two years, and references therein), as well as codes designed specifically for hardware architectures currently in development~\cite{chamberland:20,hastings:21,gidney:21}. This work represents both of these approaches, with a proof-of-principle implementation of a code designed for current IBM Quantum hardware.

In particular we consider matching codes~\cite{wootton:15}. These are the Abelian phase of Kitaev's honeycomb lattice model~\cite{honey}, reimagined as a family of stabilizer codes with similar properties to surface codes~\cite{double,dennis}. They can be defined on any trivalent lattice, but we will consider hexagonal lattices in this work.

\begin{figure}[htbp]
\begin{center}
\includegraphics[width=0.9\columnwidth]{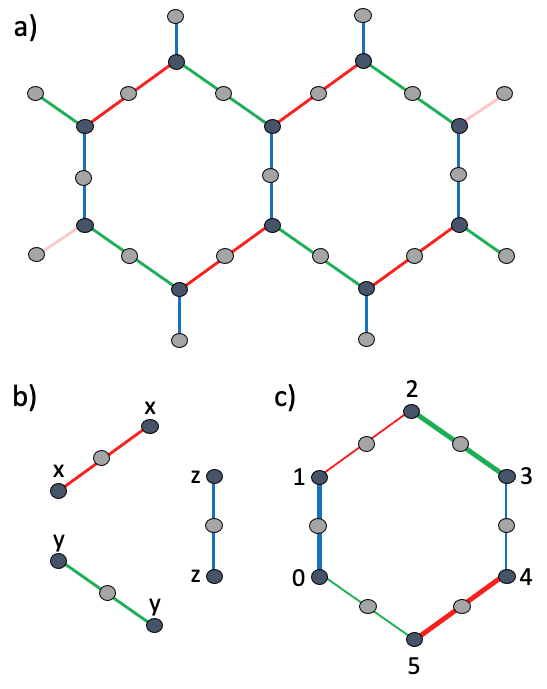}
\caption{The layout of qubits and definition of link operators in a hexagonal matching code.}
\label{code}
\end{center}
\end{figure}

Specifically, we consider a code defined on the lattice shown in Fig. \ref{code}(a). This is a so-called heavy-hexagonal lattice of qubits~\cite{chamberland:20} , in which a qubit is placed on each vertex and each edge of a hexagonal lattice. The edges of the lattice are labelled $x$, $y$ and $z$ depending on their orientation.  For each edge we define a link operator, $ \sigma^\alpha \otimes \sigma^\alpha $, that acts on the two vertex qubits. Here $\alpha \in \{x,y,z\}$ is the link type.

We also associate an operator with each plaquette. This is the product of the link operators that form the boundary of the plaquette. It therefore corresponds to the following tensor product of Paulis on the six code qubits around the plaquette.
\begin{equation}
W = \sigma_0^x \sigma_1^y \sigma_2^z \sigma_3^x \sigma_4^y \sigma_5^z
\end{equation}
Here the qubit numbering is that shown in  Fig. \ref{code}(c)

The stabilizer of the code is generated by the plaquette operators along with a commuting subset of the link operators. For concreteness, in this work we will specifically consider one such subset: that of the $z$-link operators.

Though the storage and processing of logical information is not within the scope of the current work, it is worth commenting on how this can be implemented in these codes. Logical information could be stored in the stabilizer space of the code, whose nature will depend on the boundary conditions of the lattice. However, it can also be done through careful choice of which link operators to use as stabilizers. For each pair of vertices that are not touched by a link stabilizer, a two-fold degeneracy in the stabilizer space will be opened. The code distance corresponds to the distance between these `defects'. Furthermore these defects can be moved by changing the set of link operators used as stabilizers~\cite{wootton:15,wootton:majorana}. They act as Majorana modes under braiding, allowing the implementation of some Clifford gates in a manner similar to code deformation in the surface code~\cite{bombin_twist}.

\section{Measuring stabilizers with 2-body measurements}

\begin{figure}[htbp]
\begin{center}
\includegraphics[width=0.6\columnwidth]{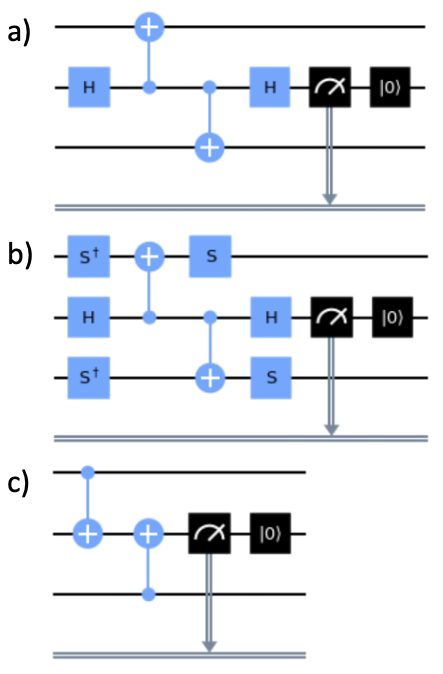}
\caption{Circuits to perform parity measurements for (a) x-links, (b) y-links and (c) z-links. The top and bottom qubits in the circuits are the pair for which the parity is measured. The middle qubit is an auxilliary, which should be initialized in the $|0\rangle$ state.}
\label{circuits}
\end{center}
\end{figure}

As with any stabilizer code, the stabilizer generators of the code must be constantly measured. Since the link operators are hermitian, they can be interpreted as observables and measured. Projections onto the eigenspaces of these operators is entangling in general, and so the measurement process requires entangling operations. The simplest means is to use an additional auxiliary qubit. This is entangled with the qubits of the link and then directly measured to give the result. The circuits required for x-, z- and y-link measurements are shown in Fig.  \ref{circuits}. The connectivity required for this circuit means that the edge qubits in the heavy-hexagonal code are perfectly placed to be the auxilliary qubit for each link.  The stabilizer generators that are link operators can therefore be directly measured in this way.

Unfortunately, the 6-body plaquette operators cannot be measured so easily. However, since the plaquette operators can be expressed as products of link operators, it is natural to expect that there could be a way to deduce their measurement from a set of link operator measurements.

The simplest way to do this is to split the 6 edges around each plaquette into two groups. In Fig. \ref{code}(c), group $a$ is shown by thin lines, and group $b$ by thick lines. The products of all link operators for each group are

\begin{eqnarray}
V_b &=& (\sigma_3^z \sigma_4^z)(\sigma_5^y \sigma_0^y)(\sigma_1^x \sigma_2^x), \nonumber \\
V_a &=& (\sigma_0^z \sigma_1^z)(\sigma_2^y \sigma_3^y)(\sigma_4^x \sigma_5^x)
\end{eqnarray}

It is important to note the following features of these operators:

\begin{enumerate}

\item They commute with the plaquette operator, which is equal to their product;
\item They commute with all individual link operators around the plaquette;
\item They commute with each other.

\end{enumerate}

From point 1 it follows that measurement of $V_a$ and $V_b$ can be used to deduce the measurement result for the plaquette operator by summing their results modulo 2. From point 2 it follows that measuring the individual link operators in a group can be used to deduce the result for a measurement of the $V$ operator of that group by summing their results modulo 2. From points 2 and 3 it follows that both $V_a$ and $V_b$ can be sequentially measured in this way, without the measurement of the latter invalidating the result of the former. Combining these, we find that the measurement of the plaquette operator can be achieved measured by measuring the link operators of first one group, and then the other, and computing the sum of all results modulo 2.

Note that this process will not commute with link stabilizers incident upon the plaquette. They will therefore need to be subsequently measured and returned to the stabilizer space. The act of measuring a plaquette operator is therefore effectively a form of code deformation~\cite{bombin:09}, which will have the effect of removing the affected stabilizer generators are removed from the stabilizer and replacing them by their product (with with the process does commute). This can lead to a decrease in the distance of the code, since it possible for errors to take shortcuts across the larger stabilizer generator without being detected. To ensure that the distance is not reduced to zero, the plaquettes will need to be measured in shifts such that sets of affected link operators do not overlap. Between the shifts, in which subsets of the plaquettes are measured, measurements of all link stabilizers are made.

A remaining question is such measurements of plaquettes are compatible with fault-tolerance. The fact that these measurements disturb z-link stabilizers means that there is effectively an additional source of noise. This is very strong (effectively applying x- and y-link operators with probability $1/2$ whenever one is measured), but also extremely localized (affecting only the measured plaquettes). Since the use of shifts means that each bout of this noise affects only non-overlapping regions, and since it is fully detected by the z-link stabilizers, a suitably calibrated decoded should be able to account for this additional noise source, with the effective loss of distance described above being the main deleterious effect.

\section{Effects of noise}

\begin{figure}[htbp]
\begin{center}
\includegraphics[width=0.8\columnwidth]{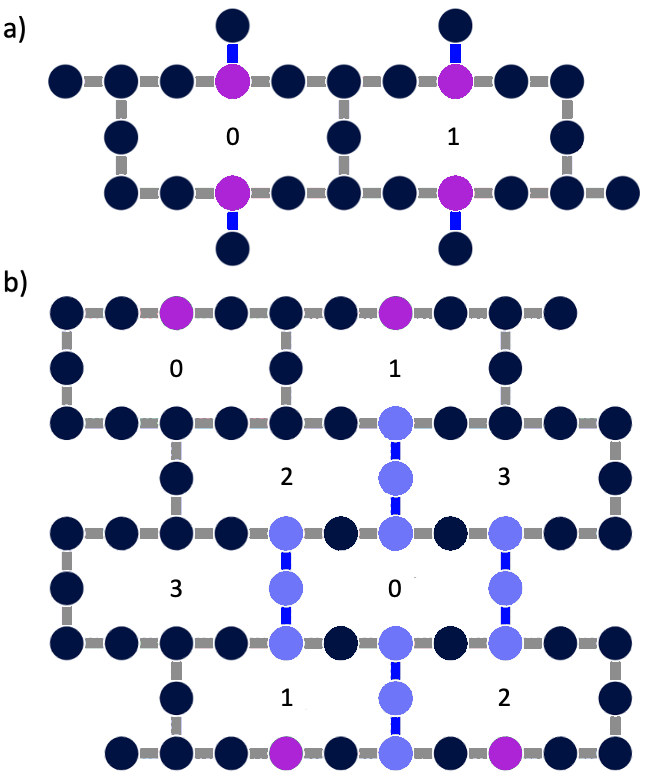}
\caption{Layout of (a) 27 qubit Falcon processors and (b) 65 qubit Hummingbird processors. The plaquettes are split into four `shifts', such that all plaquettes within each group can be measured simultaneously. Each plaquette is labelled by which shift its measurement is a part of.}
\label{devices}
\end{center}
\end{figure}

The process described above will now be implemented on multiple suitable IBM Quantum devices with the heavy-hexagon layout. These are the 27 qubit Falcon processors \texttt{ibm\_cairo}, \texttt{ibm\_hanoi}, \texttt{ibmq\_montreal} and \texttt{ibmq\_toronto}, and the 65 qubit Hummingbird processors \texttt{ibmq\_brooklyn} and \texttt{ibmq\_manhattan}. The device layouts are shown in Fig. \ref{devices}.

Note in Fig. \ref{devices} (a) that a subset of the z-links are highlighted. These are those that are incident upon plaquettes that will be measured, but for which the entire z-link is not present on the device. In these cases, the usual $ \sigma^z \otimes \sigma^z $ operator can be truncated to just a $ \sigma^z$ operator on the corresponding qubit shown in purple. The measurement of the link operator is therefore achieved by a simple measurement of this qubit. Similar truncated z-links also exist for the device in Fig. \ref{devices} (b), where the corresponding qubits are again shown in purple.

The process is repeated over $T$ measurement rounds, with $T=3$ used. Within each round are the different shifts of plaquette measurements, with each shift following by the measurement of all z-link stabilizers. Each shift consists of the measurement of all link operators within subset $a$ of the corresponding plaquettes, followed by the measurement of those from subset $b$.

When applied to the standard initial state (of all qubits in state $|0\rangle$), the plaquette operators will not have a definite initial value. The result from the first round will therefore be random for each plaquette. The results will the remain the same for subsequent rounds, except when disturbed by noise. The z-link stabilizers, however, will be affected by the measurements of x- and y-link operators. However, consider the product of all z-link operators incident upon a plaquette, as highlighted for one of the plaquettes in Fig. \ref{devices} (b). This product commutes with all the x- and y-link operators of the plaquette. The value of this product immediately before and after the measurement of the corresponding plaquette should therefore be invariant, except when disturbed by noise.

Given these considerations, we will calculate the two quantities for each plaquette, one based on the results of plaquette operator measurements and the other based on the results of z-link measurements. For the former, we consider the probability that the value of each plaquette operator remains the same as for the previous round. This done for each plaquette and each round except the first. The average over all these values, $\langle p_W \rangle$, is then calculated. Similarly we calculate a quantity $\langle p_Z \rangle$ based on the probability that the product of all z-link operators incident upon a plaquette changes values.

The process will also be run for simulated noise. This will be done for the 27 qubit case only. The noise model is the standard one used for threshold calculations of the surface code~\cite{stephens:14}: every that can go wrong does so with some a probability $p$. Specifically, bit flip noise with probability $p$ is applied after each reset and prior to each measurement, and two qubit depolarizing noise is applied to each two qubit. Note that additional noise is not applied to the single qubit gates around CX gates. Instead, the combination of these gates is regarded as a single two-qubit gate with a single error.

Note that, since the process is restricted to stabilizer states, noiseless simulations can be highly efficient. As such, the process was run for both the 27 and 65 qubit cases to verify that the calculated quantities are indeed invariant in the noiseless case.

For the runs on real hardware, regular benchmarking of the devices offers measured noise probabilities that can be compared with those of the error model. These probabilities are collected for both preparation and measurement noise on each qubit, as well as the CX noise on all pairs. Specifically, probabilities considered are as follows.
\begin{itemize}
\item Preparation noise: For each qubit, the probability \texttt{prob\_meas1\_prep0}.
\item Measurement noise: For each qubit, the measurement error probability.
\item CX noise: For each pair of qubits for which a CX is possible, the CX error probability.
\item Idle noise: For each qubit, the error probability for an identity gate over the timescale of measurement and reset. This is estimated as
$$
1-(1-2 \, p_{\rm{id}})^{(t_{\rm{meas}}+t_{\rm{reset}})/t_{\rm{id}}},
$$
Here $p_{\rm{id}}$ is the error over the standard timescale $t_{\rm{id}}$ of an identity gate. The above expression gives the probability for an odd number of such errors over the $t_{\rm{meas}}+t_{\rm{reset}}$ timescale of measurement and reset. 
\end{itemize}
The mean and standard deviation of this combined list of probabilities is shown for each device in the table below, along with the values of the quantum volume~\cite{cross:20}.

\begin{center}
\begin{tabular}{|c c c c|} 
 \hline
 Device & $\langle p \rangle$ & $\sigma$ & QV \\ [0.5ex] 
 \hline
 \texttt{ibm\_cairo} \,&\, $1.47\,\%$ \,&\, $1.23\,\%$ \,&\, 64 \\ 
 \hline
 \texttt{ibm\_hanoi} \,&\, $1.55\,\%$ \,&\, $1.61\,\%$ \,&\, 64 \\
 \hline
 \texttt{ibmq\_brooklyn} \,&\, $4.73\,\%$ \,&\, $5.00\,\%$ \,&\, 32 \\
 \hline
 \texttt{ibmq\_montreal} \,&\, $4.41\,\%$ \,&\, $6.00\,\%$ \,&\, 128 \\
  \hline
 \texttt{ibmq\_toronto} \,&\, $5.17\,\%$ \,&\, $6.45\,\%$ \,&\, 32 \\
  \hline
 \texttt{ibmq\_manhattan} \,&\, $18.3\,\%$ \,&\, $31.9\,\%$ \,&\, 32 \\
 [1ex] 
 \hline
\end{tabular}
\end{center}

Here we see that there is a high degree of variation in the probabilities, with the standard deviation on the order of $\langle p \rangle$  in all cases. We also see that there is not a strong correlation between $\langle p \rangle$ and the quantum volume. This is because the value of $\langle p \rangle$ is strongly affected by the probability of error an idle qubit while another is being measured and reset, whereas such processes do not occur in the calculation of the quantum volume. Also $\langle p \rangle$ is calculated using errors from across the whole device, whereas the quantum volume calculation can be done on an optimal subset of qubits.

\begin{figure}[htbp]
\begin{center}
\includegraphics[width=\columnwidth]{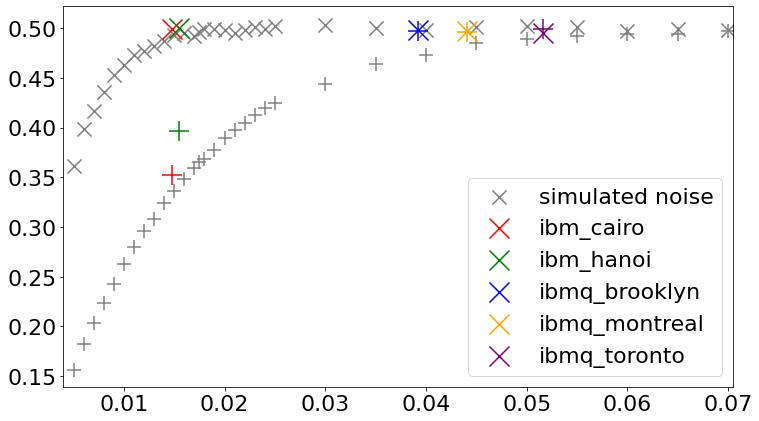}
\caption{The values $\langle p_W \rangle$ (marked as $\times$) and $\langle p_Z \rangle$ (marked as $+$) plotted against noise strength. For simulated runs, the noise parameter $p$ is used, For runs on quantum hardware, the mean value $\langle p \rangle$ is used.}
\label{results}
\end{center}
\end{figure}

The results of noisy runs, for both simulations and quantum hardware, are shown in Fig. \ref{results}. The simulations show that the probabilities $\langle p_W \rangle$ and $\langle p_Z \rangle$ increase with $p$ and converge to $\langle p_W \rangle = \langle p_Z \rangle = 0.5$, as would be expected. For any run we see that $\langle p_W$ is consistently higher than $\langle p_Z \rangle$. This will be due to the fact that there is more time between two subsequent measurements of the stabilizer operators than of the z-links, and so more errors. The value of $\langle p_W \rangle$ approaches the convergence point already at around $p=1.5\, \%$. For $\langle p_Z \rangle$, the results are easily distinguishable from the convergence point up to around $p=3\, \%$.

A further data point not shown here is for \texttt{ibmq\_manhattan}. This has $\langle p \rangle = 18.5 \%$, which is well above those for the devices shown in the graph. The values of $\langle p_W \rangle$ and $\langle p_Z \rangle$ are at the convergence point, with $\langle p_W \rangle=0.500$ and $\langle p_Z \rangle=0.498$.

For the noisiest devices tested (\texttt{ibmq\_brooklyn}, \texttt{ibmq\_montreal}, \texttt{ibmq\_toronto} and \texttt{ibmq\_manhattan}) the results show $\langle p_W \rangle \approx \langle p_Z \rangle \approx 0.5$, and so are consistent with the convergence point. This is consistent with the strength of the noise in these devices, which all have $\langle p \rangle> 4\,\%$.

For \texttt{ibm\_cairo} and \texttt{ibm\_hanoi} the results are clearly distinguishable from the convergence point, with $\langle p_Z \rangle < 0.4$ in both cases. The result for \texttt{ibm\_cairo} is in good agreement with simulations for $p \approx 1.5\%$, which is consistent with the $\langle p \rangle=1.47\,\%$ for this device. For \texttt{ibm\_hanoi} the result agrees best with simulations for $p \approx 2\%$. Though this is higher than might be expected given the $\langle p \rangle=1.55\,\%$ of this device, the high variation in the error strengths on mean that this is not a great surprise.

\section{Conclusions and Outlook}

The codes implemented in this work effectively consist of 2 independent stabilizer measurements per plaquette, per round. This means four per round for the Falcon devices, and 16 per round for the Hummingbird device. As such, these experiments are among the largest scale quantum error correction experiments yet implemented. Nevertheless, the scope of this study was not to use the codes to store or protect logical qubits. Instead it was to assess the effects of noise on the measurements of stabilizers, and to benchmark current quantum hardware by comparing results with simulations.

The results show that the various devices tested exhibit results consistent with a range of different error rates, from around $1.5\%$ in the the best case of $\texttt{ibm\_cairo}$ to $5\%$ or more for the worst case of $\texttt{ibmq\_toronto}$. The former extreme shows good progress towards the well-known threshold of $0.5\%$ -$1\%$ for the standard surface code~\cite{fowler:rev,stephens:14}. It is unfortunate that the only 65 qubit devices tested show results from the convergence point. Nevertheless, it is to be expected that initial versions of larger devices will exhibit higher error rates.

The figures of merit introduced here will continue to be applied as further quantum hardware is developed, allowing a holistic assessment of their capabilities in a concrete example of quantum error correction. Of particular interest will be the continued development of larger devices, such as the 65 qubit Hummingbird processors, to see evidence of low effective error rates. Also, the limit to $T=3$ measurement rounds in this work was due to current limitations in the classical control software. In future we hope to greatly extend this, to also assess any effect of increasing circuit depth on the error rate.

Of course, there is also the need to further develop the understanding of the code itself. This includes developing a tailored decoder, and determining the threshold of the code. This will be done in conjunction with the development of the hardware, to work towards an eventual demonstration of a logical qubit with the code.

The data for all results in this paper are available at \cite{data}

\emph{Acknowledgements:} We acknowledge support from the Swiss National Science Foundation through NCCR SPIN. 

\bibliography{references}

\end{document}